
\input phyzzx
\hoffset=0.2truein
\voffset=0.1truein
\hsize=6truein
\def\TITLEPAGE{\frontpagetrue}
\def\CALT#1{\hbox to\hsize{\tenpoint \baselineskip=12pt
        \hfil\vtop{
        \hbox{\strut hep-ph/9206216}
	\hbox{\strut CALT-68-#1}
        \hbox{\strut DOE RESEARCH AND}
        \hbox{\strut DEVELOPMENT REPORT}}}}

\def\CALTECH{
        \address{California Institute of Technology,
Pasadena, CA 91125}}
\def\TITLE#1{\vskip .5in \centerline{\fourteenpoint #1}}
\def\AUTHOR#1{\vskip .2in \centerline{#1}}

\def\ABSTRACT#1{\vskip .2in \vfil \centerline{\twelvepoint
\bf Abstract}
   #1 \vfil}
\def\ENDTITLEPAGE{\vfil\eject\pageno=1}
\hfuzz=5pt
\tolerance=10000
\TITLEPAGE
\CALT{1787}
\TITLE{Semilocal Defects\footnote\dagger{This work supported in part
by the US Department of Energy under Contract No.
DE-AC03-81-ER40050}}

\AUTHOR{John Preskill}

\CALTECH
{\baselineskip=14pt\ABSTRACT{
I analyze the interplay of gauge and global symmetries in the theory of
topological defects.
In a two-dimensional model in which both gauge symmetries and {\it exact}
global symmetries are spontaneously broken, stable vortices may fail to exist
even though magnetic flux is topologically conserved.  Following Vachaspati and
Ach\'ucarro, I formulate the condition
that must be satisfied by the pattern of symmetry breakdown for finite-energy
configurations
to exist in which the conserved magnetic flux is spread out instead of confined
to a localized vortex. If this condition is met, vortices are always unstable
at sufficiently weak gauge coupling.
I also describe the properties of defects in models with an ``accidental''
symmetry that is partially broken by gauge boson exchange.  In some cases, the
spontaneously broken accidental symmetry is not restored inside the core of the
defect. Then the
structure of the defect can be analyzed using an effective field theory; the
details of the physics responsible for the spontaneous symmetry breakdown need
not be considered.  Examples include ``semilocal'' domain walls and vortices
that are classically  unstable, but are stabilized by loop corrections, and
``semilocal'' magnetic monopoles that have an unusual core structure.  Finally,
I examine the general theory of the ``electroweak strings'' that were recently
discussed by Vachaspati.  These arise only in models with gauge boson
``mixing,'' and can always end on magnetic monopoles.  Cosmological
implications are briefly discussed.
\bigskip\rightline{June, 1992}
}}
\ENDTITLEPAGE
\eject

\chapter{Introduction}
A gauge theory that undergoes the Higgs mechanism will in many cases contain
topologically stable defects.\REF\coleman{S. Coleman, ``Classical Lumps and
Their Quantum Descendants,'' in {\it New Phenomena in Subnuclear Physics}, ed.
A. Zichichi (Plenum, London, 1977).}\REF\preskill{J. Preskill, ``Vortices and
Monopoles,'' in {\it Architecture of the Fundamental Interactions at Short
Distances}, ed. P. Ramond and R. Stora (North-Holland, Amsterdam,
1987).}\refmark{\coleman,\preskill}  For example, in two spatial dimensions,
the
classical field configurations may be classified by a conserved magnetic flux,
such that there are infinite energy barriers separating configurations with
different values of the magnetic flux.  The configuration of minimum energy in
at least one of the nontrivial magnetic flux sectors is then expected to be a
localized vortex with magnetic flux trapped in its core, a static soliton
solution to the classical field equations.
When the theory is quantized, the vortex survives as a stable particle in the
spectrum.
The corresponding defect in three spatial dimensions is a one-dimensional
string.

But it was recently noted by Vachaspati and Ach\'ucarro\Ref\vachachu{T.
Vachaspati and A. Ach\'ucarro, {\it Phys. Rev.} {\bf D44} (1991) 3067.} that,
even if magnetic flux is
topologically conserved, and a finite energy gap separates the nontrivial flux
sectors from the vacuum sector, there may be no stable vortex solutions.  This
can happen if, in addition to the spontaneously broken gauge symmetry, there is
also a spontaneously broken {\it exact} global symmetry, and so exactly
massless Nambu-Goldstone bosons in the spectrum.  The nontrivial magnetic flux
sectors may then contain configurations of finite energy in which the magnetic
flux is spread out over an arbitrarily large area, and it becomes a dynamical
question whether the energy is minimized by the localized vortex or the
configuration with unlocalized magnetic flux.  Vortices that are potentially
subject to this instability were called ``semilocal'' in Ref.~[\vachachu], in
recognition of the important role played by the global symmetry.

The purpose of this paper is to give a systematic account
of the interplay of gauge and global symmetries in the classification of
topologically stable defects,
in a more general setting than that considered in Ref.~[\vachachu].  I will
formulate
the criterion for the existence of finite energy configurations that carry a
topologically conserved magnetic flux that is unlocalized, and will note the
existence of both vortices and domain walls that are classically unstable, but
are stabilized by quantum effects involving gauge boson loops.  I will also
discuss ``semilocal monopoles'' that, while always classically stable,  can
have a different kind of core structure than the usual gauge theory monopoles.
Finally, I discuss some general properties of ``electroweak vortices,'' which
are classically stable even though they carry no topologically conserved
flux.\Ref\vach{T. Vachaspati, ``Electroweak Strings,'' Tufts Preprint TUTP-92-3
(1992); {\it Phys. Rev. Lett.} {\bf 68} (1992) 1977.}

The general approach adopted here is especially suitable for models in which
gauge or global symmetries are {\it dynamically} broken, or for any scheme in
which it is convenient to ``integrate out'' the detailed physics responsible
for the symmetry breaking.  Assuming that the relevant gauge couplings are
weak, the semilocal defects discussed here can be studied using an effective
field theory in which only light degrees of freedom are retained.  The typical
size of the defects is larger than the distance scale associated with the
symmetry breakdown by an inverse power of the weak gauge coupling.  The
distinguishing feature of ``semilocal'' defects, then, is that their detailed
structure can be analyzed without ever considering the ``restoration'' of the
spontaneously broken symmetry.

In Section 2, I describe the general class of models that will be considered in
this paper.  These models have an ``accidental'' symmetry, and part of this
symmetry is gauged.  The accidental symmetry is spontaneously broken.   To
determine the unbroken gauge group, we need to solve a ``vacuum alignment''
problem.

The general theory of semilocal vortices is discussed in Section 3.
Two cases are considered.
In the first case, there are exactly massless Nambu-Goldstone bosons, and the
topologically conserved magnetic flux need not be confined.  Stable vortices
may exist for a range of values of the gauge coupling, but vortices become
unstable when the gauge coupling is sufficiently weak.  In the second case,
there are light ``pseudo-Goldstone'' bosons; vortices become classically
unstable at weak gauge coupling, but are stabilized by quantum corrections.
Then the accidental symmetry is not restored inside the core of the vortex.

Section 4 concerns semilocal domain walls and monopoles.  These are always
stable.
They resemble the semilocal vortices in the second case above;
the core of the defect has an unusual structure, because the accidental
symmetry is not restored inside the core.

Examples that illustrate the general theory are presented in Sections 5 and 6.
The models in Section 5 have elementary Higgs fields.  The models in Section 6
do not; instead, the spontaneous symmetry breakdown is dynamical.

The criterion for the existence of configurations with unconfined magnetic flux
is further discussed in Section 7.  I show that finite-energy configurations
can exist in which the conserved magnetic flux is ``spread out'' only if gauge
and global symmetries ``mix;''  the unbroken global symmetry group must have
generators that are nontrivial linear combinations of spontaneously broken
gauge symmetry generators and global symmetry generators.

The general theory of electroweak vortices\refmark{\vach} is described in
Section 8.
These carry no conserved magnetic flux, yet are classically stable.
Their distinguishing feature is that they become stable semilocal vortices in
the limit in which some gauge coupling approaches zero.  This is
possible only if the pattern of gauge symmetry breaking admits gauge boson
mixing. (In other words, there are unbroken gauge generators that are
nontrivial linear combinations of generators that belong to distinct invariant
subalgebras of the gauged Lie algebra.)  I note that electroweak strings can
end on magnetic monopoles, and compute the magnetic charge of the monopole.  I
also discuss the Aharonov--Bohm interactions of electroweak strings, and point
out that an electroweak string cannot be used to detect the ``quantum hair'' of
an object.  Finally, I comment on the ``embedded defects'' recently discussed
by  Vachaspati and Barriola,\Ref\vachbarr{T. Vachaspati and M. Barriola, ``A
New Class of
Defects,'' Preprint (1992).} and remark that embedded monopoles are always
unstable.

Section 9 contains some concluding remarks, including comments on the
implications of electroweak strings for particle physics and cosmology.

\chapter{General Formalism}
I will consider a class of gauge theories that can be characterized as
follows:\Ref\weinberg{S. Weinberg, {\it Phys. Rev.} {\bf D13} (1976) 974.}
In the limit of vanishing gauge couplings, the theory respects a group $G_{\rm
approx}$ of global symmetries that is spontaneously broken to the subgroup
$H_{\rm approx}$.\foot{I use this notation because the $G_{\rm approx}$
symmetry
will typically be broken when the gauge interactions turn on.}  ($G_{\rm
approx}$
is a finite-dimensional compact Lie group that we will assume is connected.)
In this limit,
the theory has a degenerate vacuum manifold, and massless Nambu-Goldstone
bosons, characterized by the coset space $G_{\rm approx}/H_{\rm approx}$.

Now suppose that a subgroup $G_{\rm gauge}$ of $G_{\rm approx}$ is coupled to
gauge fields.  The gauging intrinsically breaks the $G_{\rm approx}$ symmetry
and partially lifts the vacuum degeneracy.  The surviving exact symmetry group
is the subgroup of $G_{\rm approx}$ that preserves the embedding of $G_{\rm
gauge}$ in $G_{\rm approx}$; that is,
$$
G_{\rm exact}=\{g\in G_{\rm approx}~|~g~G_{\rm gauge}~g^{-1}=
G_{\rm gauge}\}~.\eqn\exact
$$
Since $G_{\rm gauge}$ is an invariant subgroup of $G_{\rm exact}$, and $G_{\rm
exact}$ is compact, $G_{\rm
exact}$ has the local structure $G_{\rm exact}\sim G_{\rm gauge}\times G_{\rm
exact}^{\rm global}$, but it may also include discrete automorphisms of $G_{\rm
gauge}$; these will be relevant to the discussion of domain walls below.

The unbroken gauge group $H_{\rm gauge}$ is the intersection of $G_{\rm gauge}$
with $H_{\rm approx}$, and the unbroken exact symmetry group $H_{\rm exact}$ is
the intersection of $G_{\rm exact}$ with $H_{\rm approx}$.  However, these
unbroken groups cannot be determined by group theory alone.  There is, in
general, a nontrivial issue of ``vacuum alignment'' that must be resolved by
the dynamics of the theory.\refmark{\weinberg}
If we fix the embedding of $G_{\rm gauge}$ in $G_{\rm approx}$, then these
intersections depend on how $H_{\rm approx}$ is embedded in $G_{\rm approx}$;
in other words, the unbroken groups depend on how the vacuum is chosen from the
(approximate) vacuum manifold $G_{\rm approx}/H_{\rm approx}$.  The gauge
interactions lift the degeneracy of the approximate vacuum states, and
determine the alignment.  The lifting of the degeneracy is a quantum effect
arising from gauge boson loops.

Once the alignment is determined, we can divide the $G_{\rm approx}/H_{\rm
approx}$ Nambu-Goldstone bosons into three classes.  The $G_{\rm gauge}/H_{\rm
gauge}$ bosons are eaten by gauge fields.  The $G_{\rm exact}/H_{\rm exact}$
bosons that are not $G_{\rm gauge}/H_{\rm gauge}$ bosons remain exactly
massless.  And the $G_{\rm approx}/H_{\rm approx}$ bosons that are not $G_{\rm
exact}/H_{\rm exact}$ bosons acquire nonzero masses due to the gauge
interactions; they are ``pseudo-Goldstone'' bosons.\Ref\pseudo{S. Weinberg,
{\it Phys. Rev. Lett.} {\bf 29} (1972) 1698; {\it Phys. Rev.} {\bf D7} (1973)
2887.}

Though it will often be convenient to think of the breakdown of $G_{\rm
approx}$ to $H_{\rm approx}$ as due to the condensation of an elementary Higgs
scalar, the above discussion makes no assumption about the mechanism of the
symmetry breakdown.  In particular, it applies to the case of a theory that
contains no elementary scalars at all, in which the condensate is a composite
operator bilinear in elementary fermions, as in technicolor
models.\REF\susskind{S. Weinberg, {\it Phys. Rev.} {\bf D19} (1979) 1277; L
Susskind, {\it Phys. Rev.} {\bf D20} (1979) 2619.}\refmark{\weinberg,\susskind}

The symmetry breaking scheme outlined here sometimes suffers from the flaw of
``unnaturalness,'' or the need to fine-tune bare parameters.  For example, in a
theory with elementary scalars, it may be that the most general Higgs potential
of renormalizable type (a quartic polynomial in the Higgs field) that is
invariant under the $G_{\rm exact}$ symmetry is not also invariant under the
larger $G_{\rm approx}$ symmetry.  Then radiative corrections will induce
divergent symmetry breaking terms in the potential that must be removed with
suitable counterterms.  This scheme is unnatural in the sense that the feature
that the $G_{\rm approx}$ symmetry is broken {\it only} by radiative
corrections (and not by terms in the classical Higgs potential) results from a
delicate cancellation between bare parameters and radiatively induced
renormalization of parameters.

This naturalness problem is typically avoided in models without elementary
scalars, and sometimes in other cases as well.  Examples will be discussed in
Sections 5 and 6.

\chapter{Vortices}
Given the pattern of symmetry breakdown described above, let us classify the
nonsingular classical field configurations that have finite energy, in two
spatial dimensions.\refmark{\coleman}  For the Higgs field potential energy to
be finite, the
Higgs field must reside in the exact vacuum manifold $G_{\rm exact}/H_{\rm
exact}$ on the circle at $r=\infty$.  For the Higgs field gradient energy to be
finite, the Higgs field must be covariantly constant on the circle at
$r=\infty$.

Since $G_{\rm exact}$ acts transitively on the exact vacuum manifold (assuming
no exact ``accidental degeneracy''), we may perform a $G_{\rm exact}$
transformation that rotates the Higgs field at the point $(r=\infty,\theta=0)$
to a standard value $\Phi_0$.  Since it is covariantly constant, the Higgs
field on the circle at infinity must lie in the orbit of the gauge group acting
on $\Phi_0$; it can be expressed as
$$
\eqalign{\Phi(r&=\infty,\theta)=D[g(\theta)]\Phi_0~,~~g(\theta)\in G_{\rm
gauge},\cr
&g(0)=e~,\quad g(2\pi)\in H_{\rm gauge}~,\cr}\eqn\orbit
$$
where $D$ is the representation of $G_{\rm gauge}$ according to which $\Phi$
transforms.  Eq.~\orbit\
associates with each finite energy field configuration a  closed path in the
coset space $G_{\rm gauge}/H_{\rm gauge}$ that begins and ends at the trivial
coset.  Thus, the nonsingular field configurations of finite energy can be
classified by the fundamental group $\pi_1(G_{\rm gauge}/H_{\rm gauge})$.
\foot{Actually, there is an ambiguity in this correspondence when $H_{\rm
gauge}$ is nonabelian and disconnected. This ambiguity can be resolved if we
consider patching together distantly separated configurations; it has no effect
on the ensuing discussion.}  There is an infinite energy barrier separating
configurations that correspond to different elements of this group, while
configurations that correspond to the same element can be smoothly deformed one
to another, while the energy remains finite.

Because $G_{\rm gauge}\subseteq G_{\rm exact}\subseteq G_{\rm approx}$, we have
the inclusion $G_{\rm gauge}/H_{\rm gauge}\subseteq G_{\rm exact}/H_{\rm
exact}\subseteq G_{\rm approx}/H_{\rm approx}$ (for the coset spaces are
obtained by the action of the groups on $\Phi_0$).  Thus, there are natural
homomorphisms
$$\eqalign{
&\pi_1(G_{\rm gauge}/H_{\rm gauge})\longrightarrow
\pi_1(G_{\rm exact}/H_{\rm exact})~, ~~~\cr
&\pi_1(G_{\rm exact}/H_{\rm exact})\longrightarrow
\pi_1(G_{\rm approx}/H_{\rm approx})~;}
\eqn\homos
$$
each loop in $G_{\rm gauge}/H_{\rm gauge}$ is also a loop in $G_{\rm
exact}/H_{\rm exact}$ and each loop in $G_{\rm exact}/H_{\rm exact}$ is also a
loop in $G_{\rm approx}/H_{\rm approx}$.

We can distinguish three types of elements of $\pi_1(G_{\rm gauge}/H_{\rm
gauge})$, according to whether the element belongs to the kernel of these
homomorphisms.  First consider an element that is not in the kernel of either
homomorphism.  This means that the corresponding noncontractible loop in the
gauged vacuum manifold $G_{\rm gauge}/H_{\rm gauge}$ remains noncontractible in
the (larger) approximate vacuum manifold.  Hence, the finite energy field
configurations associated with this loop cannot lie in the approximate vacuum
manifold everywhere.  Each field configuration must therefore have a ``core''
somewhere where the Higgs field potential energy density is nonvanishing.  If
we minimize the energy in this topological sector, we will obtain a static
vortex solution to the classical field equations, or perhaps a configuration of
two or more widely separated vortices.

Second, consider a nontrivial element of $\pi_1(G_{\rm gauge}/H_{\rm gauge})$
that is in the kernel of the first homomorphism.  This means that the
corresponding noncontractible loop in $G_{\rm gauge}/H_{\rm gauge}$ can be
contracted in the exact vacuum manifold $G_{\rm exact}/H_{\rm exact}$.  Hence,
we can construct finite energy configurations in this class that live in the
exact vacuum manifold everywhere, and have no Higgs field potential energy.

(Because $G_{\rm exact}$ has the local structure $G_{\rm exact}\sim G_{\rm
gauge}\times G_{\rm exact}^{\rm global}$, this kernel can be nontrivial only if
there is mixing of gauge and global symmetries.  That is, $H_{\rm exact}$ must
be nontrivial, and there must be $H_{\rm exact}$ generators that are  linear
combinations of $G_{\rm gauge}$ generators and $G_{\rm exact}^{\rm global}$
generators.  I will discuss this point further in Section 7.)

To understand these configurations better, consider the classical field theory
in the limit of {\it infinite} gauge coupling.  Then the gauge field is
nondynamical---gauge fields carry no energy.  Still the gauging has nontrivial
consequences, for Higgs field configurations that differ by a gauge
transformation are effectively identified.  The {\it physical} vacuum manifold
is not $G_{\rm exact}/H_{\rm exact}$, but rather this coset space with the
action of the gauge group $G_{\rm gauge}$ modded out.  That is, it is the space
$M_{\rm orbit}$ of $G_{\rm gauge}$ orbits on $G_{\rm exact}/H_{\rm exact}$.

In this limit, the configurations such that the Higgs field lies in the exact
vacuum manifold everywhere have only gradient energy.  And gradient energy in
two spatial dimensions is scale invariant.  Thus, if we find the configuration
of this type that has minimal energy, there will actually be an infinite set of
such configurations, parametrized by an arbitrary size scale.  What we have
constructed is a two dimensional ``skyrmion''\Ref\skyrme{T.~H.~R.~Skyrme, {\it
Proc. Roy. Soc.} {\bf A260} (1961) 127; E. Witten, {\it Nucl. Phys.} {\bf B223}
(1983) 433.} (or ``global texture''\Ref\kibble{T.~W.~B.~Kibble, {\it J. Phys.}
{\bf A9} (1976) 1387.}) associated with a topologically
nontrivial mapping from the two-sphere (the plane plus the point at infinity)
to the physical vacuum manifold $M_{\rm orbit}$.  Its energy
will be
$$m_{\rm skyrmion}=
Cv^2~,
\eqn\skyrmassC
$$
where $v$ is the symmetry breaking scale and $C$ is a numerical constant of
order one, for $v$ is the only relevant scale.

Now let us re-introduce the gauge field kinetic term.  The skyrmions that we
have constructed carry nonzero magnetic flux.  (The gauge field cannot be a
pure gauge everywhere, because it is topologically nontrivial on the circle at
infinity, and is smooth on the plane.)  When the gauge field dynamics turns on,
this flux will want to spread out.  The skyrmion of infinite size now will have
the lowest energy; in fact, its gauge field energy will vanish.

What we have found, then, is that in a sector whose ``magnetic flux'' is
characterized by a noncontractible loop in $G_{\rm gauge}/H_{\rm gauge}$ that
can be contracted in $G_{\rm exact}/H_{\rm exact}$, configurations of finite
energy can be constructed such that the flux is spread out over an arbitrarily
large area.  This sector is separated from sectors with other values of the
flux by an infinite energy barrier.  But within this sector there are
configurations in which the energy density is arbitrarily small everywhere
(although the total energy is bounded from below by $Cv^2$).  Notice that this
is possible only in a theory that contains exactly massless Nambu-Goldstone
bosons, for only then can a scale-invariant skyrmion exist.

In a magnetic flux sector of this type, there will of course also be
configurations in which the magnetic flux is trapped inside a vortex core where
the Higgs field leaves the exact vacuum manifold.  It becomes a dynamical
question (not a topological one) whether the vortex configurations or the
spread out configurations have lower energy.  In the limit of large gauge
coupling, where magnetic field energy can be ignored, the vortex energy is of
order $v^2$.  (This follows from dimensional analysis, assuming that there are
not large dimensionless {\it ratios} of parameters in the Higgs potential.)
Then, whether a vortex is stable or not presumably depends on the details of
the Higgs potential.  But a definite statement can be made about the opposite
limit of weak gauge coupling.  In this limit, the vortex carries enormous
magnetic flux that must spread out.  Any configuration with a Higgs field core
that remains bounded in this limit carries an energy that scales like
$$
m_{\rm vortex}\sim v^2\log(1/e^2)~,
\eqn\vortex
$$
where $e$ is the gauge coupling.  This behavior results from the competition
between Higgs field gradient energy of order $v^2\log(r)$ and magnetic field
energy of order $1/(e^2 r^2)$, where $r$ is the size of the region occupied by
the the flux.  Thus, when the gauge coupling is sufficiently weak, the skyrmion
configuration minimizes the energy, and there is no stable vortex in this flux
sector.

Even if the skyrmion minimizes the energy in a magnetic flux sector, there may
be a vortex configuration (with finite core size) in the same sector that is
classically stable.  The vortex will then be metastable and will decay via
quantum tunneling.  From the Euclidean path integral viewpoint, the instanton
configuration that mediates the decay is a ``global
monopole.''{}\Ref\vilemono{M.
Barriola and A. Vilenkin, {\it Phys. Rev. Lett.} {\bf 63} (1989) 314.}  In the
limit
of infinite gauge coupling, this is a configuration with a nontrivial Higgs
field core, where the Higgs field on a large sphere surrounding the core
assumes the nontrivial mapping from the two-sphere to the exact vacuum manifold
that is associated with the skyrmion.  For finite gauge field coupling, this
configuration has magnetic flux that enters the core from a narrow tube (the
vortex) and then  spreads out and returns to infinity (the skyrmion).
Similarly, a string in three spatial dimensions is metastable for this range of
parameters, because the string can break by nucleating a global
monopole-antimonopole pair.  The long-range interaction energy between a pair
of global monopoles with separation $r$ is $Cv^2 r$ (with $C$ defined by
eq.~\skyrmassC), so it is energetically
favorable for the monopole pair to form if the string tension is greater than
$Cv^2$.\REF\presvile{J. Preskill and A. Vilenkin, ``Decay of Metastable
Topological Defects,'' Harvard Preprint HUTP-92/A018 (1992).}  These decay
processes are further discussed in Ref.~[\presvile].

Finally, consider an element of $\pi_1(G_{\rm gauge}/H_{\rm gauge})$ that is in
the kernel of the second homomorphism in eq.~\homos\ but not the first.  This
means that the corresponding noncontractible loop in $G_{\rm gauge}/H_{\rm
gauge}$ remains noncontractible in $G_{\rm exact}/H_{\rm exact}$, but can be
contracted in $G_{\rm approx}/H_{\rm approx}$.  Hence, we can construct
configurations in this flux sector such that the Higgs field lies in the
approximate vacuum manifold everywhere, but not configurations  that lie in the
exact vacuum manifold everywhere.  When the gauge coupling is sufficiently
weak,
the vortex solutions become classically unstable, and the flux wants to spread
out.  But quantum corrections due to gauge boson exchange prevent the vortex
from spreading to infinity.

\chapter{Domain Walls and Monopoles}
Within the symmetry breaking scheme formulated in Section 2, we may also
consider the properties of topological domain walls and monopoles.  Though
there are no unexpected instabilities, these defects can have some unusual
properties that are worthy of note.

\section{Domain Walls}
The nonsingular configurations that have finite energy in one spatial dimension
are classified by the group $\pi_0(G_{\rm exact}/H_{\rm exact})$.\foot{We need
not be concerned with the ambiguity in this classification that can arise when
$H_{\rm exact}$ is nonabelian.}
For the Higgs field potential energy to be finite, the Higgs field must take a
value in the exact vacuum manifold $G_{\rm exact}/H_{\rm exact}$ at both points
at infinity.  By performing a suitable $G_{\rm exact}$ transformation, we may
choose the Higgs field at $x=-\infty$ to assume the standard value $\Phi_0$.
Two configurations can be smoothly deformed one to the other while the energy
remains finite if and only if $\Phi(x=\infty)$ for both configurations lies in
the same connected component of the exact vacuum manifold.  By minimizing the
energy in a nontrivial sector, we construct a static domain wall solution to
the classical field equations (or perhaps two or more distantly separated
domain walls).

A nontrivial element of $\pi_0(G_{\rm exact}/H_{\rm exact})$ may be in the
kernel of the homomorphism
$$
\pi_0(G_{\rm exact}/H_{\rm exact})\longrightarrow
\pi_0(G_{\rm approx}/H_{\rm approx})~;
\eqn\wallhomo
$$
that is, a vacuum state that is not connected to $\Phi_0$ in the exact vacuum
manifold may be connected to $\Phi_0$ in the approximate vacuum manifold.  Then
the domain wall will be classically unstable.  It can be deformed to a
configuration that has no classical Higgs potential energy, and it will then
want to spread out to minimize its gradient energy.  But the quantum
corrections to the effective Higgs potential, generated by gauge boson
exchange, will prevent the domain wall from spreading indefinitely, and will
stabilize it.

\section{Monopoles}
In order that a field configuration  have finite energy in three spatial
dimensions, the Higgs field must takes values in $G_{\rm exact}/H_{\rm exact}$
on the two-sphere at $r=\infty$, and must be covariantly constant on the
two-sphere.  Thus, nonsingular finite energy configurations are classified
by\refmark{\coleman}
$\pi_2(G_{\rm gauge}/H_{\rm gauge})=\pi_1(H_{\rm gauge})/\pi_1(G_{\rm
gauge})$;\foot{We need not be concerned with the ambiguity in this
classification that can arise when $H_{\rm gauge}$ is disconnected and
nonabelian.}
they are associated with noncontractible closed paths in $H_{\rm gauge}$,
beginning and ending at the identity, that
are contractible in $G_{\rm gauge}$.  The element of $\pi_1(H_{\rm gauge})$
associated with a nontrivial sector identifies the topologically conserved
magnetic charge of that sector.\REF\lubkin{E. Lubkin, {\it Ann. Phys.} {\bf 23}
(1963) 233.}\refmark{\lubkin,\coleman}

An element of $\pi_2(G_{\rm gauge}/H_{\rm gauge})$ is in the kernel of the
homomorphism
$$
\pi_2(G_{\rm gauge}/H_{\rm gauge})\longrightarrow
\pi_2(G_{\rm approx}/H_{\rm approx})
\eqn\monohomo
$$
if the noncontractible loop in $H_{\rm gauge}$ is contractible in $H_{\rm
approx}$.\foot{Note that it is not possible for this loop to be contractible in
$H_{\rm exact}$.  This is because $H_{\rm exact}$ has the general form $H_{\rm
exact}=[H_{\rm gauge}\times H_{\rm exact}^{\rm global}]/H_{\rm discrete}$,
where $H_{\rm discrete}$ is a discrete invariant subgroup of $H_{\rm
gauge}\times H_{\rm exact}^{\rm global}$.}  A sector with
this property has nontrivial magnetic charge, but also contains configurations
such that the Higgs field lies in the approximate vacuum manifold everywhere.
Ignoring quantum effects, these configurations have only gradient energy and
magnetic field energy.  The gradient energy makes them want to shrink, but they
are prevented from collapsing completely by their magnetic field energy.

These ``semilocal'' magnetic monopoles have a different core structure than the
usual gauge
theory monopoles.\REF\gibbons{G. W. Gibbons, M. E. Ortiz, F. Ruiz Ruiz, and T.
M. Samols, ``Semilocal Strings and Monopoles,'' Cambridge Preprint CTP\#2063
(1992).}\foot{The term ``semilocal monopole'' is used differently here than in
Ref.~[\vachachu] and [\gibbons].}  ``Heavy'' broken gauge fields are excited in
the core, and
the embedding of $G_{\rm gauge}$ in $G_{\rm approx}$ varies in the core, but
the spontaneously broken $G_{\rm approx}$ symmetry is not ``restored''
anywhere.  It is a dynamical question, depending on the details of the Higgs
potential, whether the realization of the $G_{\rm approx}$ symmetry actually
changes inside the core of the monopole configuration with minimal energy.

I should clarify the difference between  semilocal monopoles and the monopoles
that arise in typical grand unified theories.
It is a general feature, shared by semilocal monopoles and monopoles of the
usual kind, that the realization of the {\it gauge} symmetry must be different
inside the monopole core than in the vacuum (at least at an isolated point
inside the core).  This is not to say that the gauge symmetry is fully restored
inside the core.
In the $SU(5)$ model, for example, if we ignore the electroweak symmetry
breakdown, a Higgs field in the adjoint representation breaks the gauge
symmetry to $[SU(3)\times SU(2)\times U(1)]/Z_6$.  Inside the core of the
minimally charged magnetic monopole, the stability group of the Higgs field is
reduced to a subgroup of the symmetry of the vacuum;\Ref\dokos{C. Dokos and T.
Tomaras, {\it Phys. Rev.} {\bf D21} (1980) 2940.} namely, $[SU(2)\times
U(1)\times U(1)\times U(1)]/[Z_6\times Z_2]$.  At the center of the core, this
symmetry is enhanced to $[SU(2)\times SU(2)\times U(1)\times U(1)]/[Z_2\times
Z_2]$.

This example illustrates the generic case.  The symmetry $H_{\rm core}$
inside the core is a subgroup of the symmetry $H_{\rm gauge}$ in the vacuum.
The topological magnetic charge of the monopole can be characterized by a
noncontractible closed path in $H_{\rm core}$ that begins and ends at the
identity.  In order for the Higgs field to be smooth, this symmetry must
enlarge at the center of the core to $H_{\rm center}\supset H_{\rm core}$, such
that this closed path in $H_{\rm core}$ can be contracted in $H_{\rm center}$.
We see that
$H_{\rm center}$ cannot be contained in $H_{\rm gauge}$, but it is not
necessary for $H_{\rm center}$ to contain $H_{\rm gauge}$, either.

In a semilocal monopole, too, the subgroup $H_{\rm center}$ of $G_{\rm gauge}$
that preserves the Higgs field at the center of the core is not contained in
$H_{\rm gauge}$.
But this is achieved even though the stability group of the Higgs field is
$H_{\rm approx}$ everywhere.  The realization of the gauge symmetry changes
inside the core because the relative alignment of $H_{\rm approx}$ and $G_{\rm
gauge}$ adjusts there.  This means that the core of the monopole can be
accurately described in an ``effective field theory'' that describes physics
below the scale of the symmetry breakdown, as I will discuss in more detail in
Section 6.

\chapter{Examples}
I will now apply the above discussion to a sample model.  When all gauge
interactions are turned off, the Lagrange density of this model is
$$
{\cal L}={1\over 2}\partial_{\mu}\pi^a\partial^{\mu}\pi^a
-{\lambda\over8}\left(\pi^a\pi^a-v^2\right)^2~,
\eqn\lagrangian
$$
where $a=0,1,2,3$.  Thus, in this limit, the symmetry breaking pattern of the
model is
$$
G_{\rm approx}=SO(4)\to H_{\rm approx}=SO(3)~,
\eqn\pattern
$$
and there are three Nambu-Goldstone bosons, plus one massive Higgs field with
mass
$$
m^2_S=\lambda v^2~.
\eqn\higgsmass
$$
It is convenient to write the Higgs field as a two-by-two matrix
$$
\Phi={1\over\sqrt{2}}\left(\pi^0+ i\vec{\pi}\cdot\vec{\sigma}\right)
\equiv\left(\matrix{\phi_1&-\phi_2^*\cr\phi_2&\phi_1^*\cr}\right)~,
\eqn\mat
$$
which transforms under $G_{\rm approx}=[SU(2)_L\times SU(2)_R]/Z_2$ according
to
$$
\Phi\to U_L\Phi U_R^\dagger~.
\eqn\transform
$$

\section{Unstable Vortex}
Let us briefly recall the model that was analyzed in Ref.~[\vachachu].  It is
obtained
by gauging the $U(1)$ subgroup of one of the $SU(2)$'s.  We choose to gauge the
$U(1)_R$ generated by
$$
Y_R={\rm diag}\left({1\over 2}~,-{1\over 2}\right)~.
\eqn\hyper
$$
Then the exact symmetry of the model is
$$
G_{\rm exact}=\left(SU(2)_L\times \left[U(1)_R\times_{SD} Z_{2,R}
\right]\right)/ Z_2~,
\eqn\exactA
$$
where $\times_{SD}$ denotes a semi-direct product.  Here the $Z_{2,R}$ is
generated by the charge conjugation operation
$$
{\cal C}_R:\phi\equiv \pmatrix{\phi_1\cr \phi_2\cr}
\to \phi^c=-i \sigma_2\phi^*=
\pmatrix{-\phi_2^*\cr \phi_1*\cr}~.
\eqn\conjugation
$$
This operation commutes with $SU(2)_L$, but anticommutes with $Y_R$,
$$
{\cal C}_R Y_R{\cal C}_R^{-1}=-Y_R~;
\eqn\anticommute
$$
it is a nontrivial automorphism of the $U(1)_R$ gauge group.  In this case, the
$G_{\rm approx}$ symmetry is ``natural,'' because the potential in
eq.~\lagrangian\ is the most general quartic potential with the $G_{\rm exact}$
symmetry.

Here $G_{\rm exact}$ acts transitively on $G_{\rm approx}/H_{\rm approx}$, so
the alignment problem is trivial.  Any Higgs field in the $G_{\rm
approx}/H_{\rm approx}$ can be rotated by a $G_{\rm exact}$ transformation to
the standard form
$$
\phi_0={1\over\sqrt{2}}\pmatrix{v\cr 0\cr}~.
\eqn\standard
$$
The gauge symmetry is completely broken, and the unbroken exact symmetry is
$$
H_{\rm exact}=U(1)_V\times_{SD} Z_{2,V}~,
\eqn\exactvec
$$
where $U(1)_V$ is generated by
$$
Q=Y_L + Y_R
\eqn\qdef
$$
and the $Z_{2,V}$ is generated by the charge conjugation operation
$$
{\cal C}_V:\phi\to \phi^*
\eqn\veccc
$$
that anticommutes with $Q$.
Of the three Nambu-Goldstone bosons, one is eaten, and the other two remain
exactly massless.  The $U(1)_R$ vector boson acquires the mass
$$
\mu^2={1\over 4}e^2 v^2~.
\eqn\vectormass
$$

This model has no stable domain walls or monopoles, but it has a topologically
conserved magnetic flux classified by $\pi_1(U(1)_R)=Z$.  The vortex
configuration with unit flux has the asymptotic behavior
$$
\phi(r=\infty,\theta)={v\over\sqrt{2}}\pmatrix{e^{i\theta}\cr 0\cr}~.
\eqn\vorasym
$$
Since $\pi_1(G_{\rm exact}/H_{\rm exact})=0$, the first homomorphism in
eq.~\homos\ has trivial kernel, and a cylindrically symmetric ``skyrmion''
configuration can be constructed that has this asymptotic behavior, and lives
in the exact vacuum manifold everywhere; it is
$$
\phi^{\rm (skyrmion)}(r,\theta)={v\over\sqrt{2}}(r^2+a^2)^{-{1\over 2}}
\pmatrix{re^{i\theta}\cr a\cr}~,\quad
{e\over 2}A^{\rm (skyrmion)}_{\theta}(r)={-r^2\over r^2+a^2}~,
\eqn\skyrsol
$$
where $a$ is an arbitrary distance scale, and $e$ is the gauge coupling.  As
Hindmarsh\Ref\hindmarsh{M. Hindmarsh, {\it Phys. Rev. Lett.} {\bf 68} (1992)
1263.} observes (see also Ref.[\gibbons]), the exact vacuum manifold, with the
gauged $U(1)_R$ modded
out, is the manifold $CP^1=S^2$, and eq.~\skyrsol\ is the skyrmion solution of
the $CP^1$ sigma model in two spatial dimensions.  Its covariant gradient
energy (in two
dimensions) is
$$
m_{\rm skyrmion}=\pi v^2~,
\eqn\skyrene
$$
which is thus the energy of the configuration with the magnetic flux spread out
to infinity.

There is also a Nielsen-Olesen\Ref\nielsen{H. B. Nielsen and P. Olesen, {\it
Nucl. Phys.} {\bf B61} (1973) 45.} vortex solution, with $\phi=0$ at the
origin.
Its mass equals the skyrmion mass for
$\beta\equiv\lambda/(e/2)^2=m_S^2/\mu^2=1$, and it is
lighter than the skyrmion for $\beta<1$.\Ref\bogo{E. Bogomol'nyi, {\it Sov. J.
Nucl. Phys.} {\bf 24}
(1976) 449.}  Thus, there is a stable vortex for
$\beta<1$.  But for $\beta>1$, the Nielsen-Olesen vortex solution is heavier
than the
skyrmion, and the vortex is unstable.  The analysis of
Hindmarsh\refmark{\hindmarsh} and of Ach\'ucarro and
Vachaspati\Ref\achuvach{A. Ach\'ucarro, K. Kuijen, L. Perivolaropoulos,  and T.
Vachaspati, ``Dynamical
Simulations of Semilocal Strings,'' CFA Preprint 3384 (1992).} indicates that
there are no metastable vortices in this model.  For
$\beta>1$, the vortex is classically unstable, and the magnetic flux wants to
spread out.

\section{Quantum Stability}
Now consider gauging
$$
G_{\rm gauge}=[U(1)_L\times U(1)_R]/Z_2
\eqn\newgauge
$$
generated by $Y_L$ and $Y_R$.  The exact symmetry of this model is
$$
G_{\rm exact}=([U(1)_L\times_{SD} Z_{2,L}]
\times [U(1)_R\times_{SD} Z_{2,R}])/Z_2
\eqn\newexact
$$
where the $Z_2$'s are generated by the charge conjugation operations
$$
\eqalign
{&{\cal C}_L: \phi\to -i\sigma_2 \phi~,\cr
&{\cal C}_R: \phi\to -i\sigma_2 \phi^*~;\cr}
\eqn\newcc
$$
${\cal C}_L$ flips the sign of $Y_L$, and ${\cal C}_R$ flips the sign of $Y_R$.

Unfortunately, the $G_{\rm approx}$ symmetry is not natural in this model. The
quartic interaction term $(|\phi_1|^2-|\phi_2|^2)^2$ is invariant under $G_{\rm
exact}$, but has not been included in eq.~\lagrangian.  I will nevertheless
analyze the effect of symmetry-breaking quantum corrections in this model, to
illustrate the earlier general discussion.  Natural models can be constructed
(notably including models without elementary scalars, in which the spontaneous
breakdown of $G_{\rm approx}$ is {\it dynamical}), but they are more
complicated to construct and analyze.  Examples will be discussed in Section 6.

There is a nontrivial alignment problem in this model, which we can resolve by
minimizing the one-loop effective potential.  If the Higgs doublet has the
vacuum expectation value
$$
\VEV{\phi}\equiv \phi_0={1\over \sqrt{2}}\pmatrix{v_1\cr v_2\cr}~,
{}~~|v_1|^2+|v_2|^2=v^2~,
\eqn\vacval
$$
then the tree-level gauge boson mass matrix is
$$
\mu^2={1\over 4}\pmatrix{e_L^2(|v_1|^2+|v_2|^2)&e_L e_R(-|v_1|^2+|v_2|^2)\cr
e_L e_R(-|v_1|^2+|v_2|^2)&e_R^2(|v_1|^2+|v_2|^2)\cr}~,
\eqn\massmat
$$
where $e_{L,R}$ are the gauge couplings, and the leading (in $\hbar$) term in
the effective potential that depends on the alignment is\Ref\colewein{S.
Coleman and E. Weinberg, {\it Phys. Rev.} {\bf D7} (1973) 1888.}
$$
V_{\rm eff}={3\over 64\pi^2}\Tr \mu^4\ln(\mu^2/M^2)~,
\eqn\effpot
$$
where $M$ is a renormalization scale.

A symptom of the unnaturalness of this model is that the statement that the
$G_{\rm approx}$ symmetry is a good symmetry at the classical level is really
dependent on the choice of the scale $M$, for shifting the renormalization
scale moves symmetry-breaking terms in the potential from the one-loop term to
the tree term.  We suppose that the classical potential is $G_{\rm
approx}$-invariant when $M$ is of order the symmetry breaking scale $v$.  Then
the minimum of the potential occurs for $|v_1|^2|v_2|^2=0$, if the gauge
couplings are weak.  By a $G_{\rm exact}$ transformation, we can therefore
choose
$\phi_0$ as in eq.~\standard.  Thus the unbroken symmetries are
$$
H_{\rm gauge}=U(1)_V~, \quad H_{\rm exact}=U(1)_V\times_{SD} Z_{2,V}~,
\eqn\vectorunbroken
$$
with the $U(1)_V$ and $Z_{2,V}$ generators defined as in eq.~\qdef-\veccc.
The vector boson spectrum is
$$
\eqalign{
&\mu^2_Z={1\over 4}(e_L^2+e_R^2)v^2~,\quad Z= B_L\cos\theta  - B_R\sin\theta
{}~,\cr
&\mu^2_A=0~,\quad\quad\quad\quad\quad\quad~ A= B_L\sin\theta +B_R\cos\theta
{}~,\cr}
\eqn\bosspec
$$
where $B_L$ and $B_R$ are the $U(1)_L\times U(1)_R$ gauge bosons, and $\theta$
is the mixing angle defined by $\tan\theta=e_R/e_L$.
Of the three Nambu-Goldstone bosons, one is eaten, and the other two become a
charged pseudo-Goldstone boson with mass
$$
m^2_{\rm PGB}=-~{3\over 128\pi^2}e_L^2 e_R^2 v^2\ln[(e_L^2+e_R^2)v^2/4M^2]~;
\eqn\pgb
$$
the mass goes to zero as either gauge coupling turns off.

Since there is a spontaneously broken $U(1)$ gauge symmetry, there is a
topologically conserved magnetic flux, and vortex configurations can be
constructed that have $Z$ flux trapped in the core.  If we ignore the quantum
corrections to the effective potential, the stability of the vortices can be
analyzed just as above, except that the critical coupling becomes
$\beta=\lambda/[(e_L^2+e_R^2)/4]=m_S^2/\mu_Z^2=1$.  For $\beta>1$, the vortex
becomes
classically unstable, and the magnetic flux wants to spread.  But in this
model, the one-loop corrections prevent the flux from spreading to infinity.
The vortex, stabilized by quantum corrections, has a core size
$$
r_{\rm core}\sim m_{\rm PGB}^{-1}~.
\eqn\coresize
$$

This model also contains domain walls, because the exact vacuum manifold has
two disconnected components---one with $|v_1|^2=0$ and one with $|v_2|^2=0$.
Ignoring quantum corrections, the domain wall is unstable; it can lower its
gradient energy by spreading out.  But the one-loop corrections prevent it from
spreading beyond a size given by eq.~\coresize.

\section{Semilocal Monopole}
Now suppose that the gauge group is
$$
G_{\rm gauge}=SO(3)~,
\eqn\monogauge
$$
under which $\pi^a,~a=1,2,3$, is a triplet and $\pi^0$ is a singlet.  Then the
exact symmetry is
$$
G_{\rm exact}=O(3)~
\eqn\monoexact
$$
(which includes a parity transformation---the element $-{\bf 1}$ in $SO(4)$).

Again, this model is unnatural; we can add any even function of $\pi^0$ to the
potential in eq.~\lagrangian\ without breaking the $G_{\rm exact}$ symmetry.
Still, we may proceed as in Section 5.2, imposing the symmetry at a
renormalization scale of order $v$, and solving for the alignment by minimizing
the one-loop effective potential.  We then find that the minimum occurs for
$\pi^0=0$, so that the unbroken gauge symmetry is
$$
H_{\rm gauge}=SO(2)~.
\eqn\monogaex
$$
(For a caveat concerning this alignment, see the discussion of dynamical
symmetry breaking in Section 6)  Thus, $\pi_2(G_{\rm gauge}/H_{\rm
gauge})=Z$, and this model contains magnetic monopoles.

Since $\pi_2(G_{\rm approx}/H_{\rm approx})=0$, the homomorphism eq.~\monohomo\
has a trivial kernel.  This means that there are magnetically charged
configurations such that the Higgs field takes values in the approximate vacuum
manifold everywhere.  A spherically symmetric configuration of this type that
carries one unit of magnetic charge is
$$
\eqalign{
&\pi^0=v\sqrt{1-f(r)^2}~,\cr
&\pi^a=vf(r)\hat r^a,\quad a=1,2,3~,\cr}
\eqn\hedgehog
$$
where
$$
f(\infty)=1~,\quad f(0)=0~.
\eqn\fasym
$$
This configuration can lower its gradient energy by shrinking, but
it is prevented from collapsing completely by its magnetic Coulomb energy.  If
$\pi$ is constrained to take the form eq.~\hedgehog, then the energy will be
minimized when the size is of order $(ev)^{-1}$ (where $e$ is the gauge
coupling), and the mass of the monopole is of order $4\pi v/e$.

There are also ``'t Hooft--Polyakov'' configurations\Ref\hooft{G. 't Hooft,
{\it Nucl. Phys.} {\bf B79} (1974) 276; A. M. Polyakov, {\it JETP Lett.} {\bf
20} (1974) 194.}
$$
\eqalign{
&\pi^0=0~,\cr
&\pi^a=v g(r) \hat r^a~,a=1,2,3~,\cr}
\eqn\hphedge
$$
where
$$
g(\infty)=1~,\quad g(0)=0~.
\eqn\hpasym
$$
Such a configuration has Higgs field potential energy in its core, and the
energy is minimized by the usual 't Hooft-Polyakov solution.

In the Bogomol'nyi limit $\lambda/e^2\to 0$,\REF\prasad{M. Prasad and C.
Sommerfield, {\it Phys. Rev. Lett.} {\bf 35} (1975) 760; S. Coleman, S. Parke,
A. Neveu, and C. Sommerfield, {\it Phys. Rev.} {\bf D15} (1977)
544.}\refmark{\bogo,\prasad} the Higgs potential energy is
negligible, and the monopole of minimal energy has the form eq.~\hphedge.
Turning on $\pi^0$ only increases the gradient energy.  But in the opposite
limit $\lambda/e^2\to \infty$, the form eq.~\hedgehog\ has lower energy.  To
see
this, note that in the limit of large $\lambda$, the Higgs field core of the 't
Hooft--Polyakov solution shrinks to zero size,\Ref\zachos{T. W. Kirkman and C.
K. Zachos, {\it Phys. Rev.} {\bf D24} (1981) 999.} so that $g(r)=0$, for all
$r$.
This solution is then of the form eq.~\hedgehog, but with $f(r)$ constrained to
be 1.  Evidently, by relaxing this constraint, a lower energy configuration of
the form eq.~\hedgehog\ can be found.  Thus, for large $\lambda$, the Higgs
field inside the monopole core remains close to the approximate vacuum
manifold, and the approximate $SO(4)$ symmetry is not ``restored'' anywhere
inside the core.  This is a semilocal monopole.

A {\it natural} model with a semilocal monopole can be constructed as
follows:
Consider the symmetry breaking pattern $G_{\rm approx}=SO(8)\to H_{\rm
approx}=SO(7)$, driven by a Higgs field in the vector representation of
$SO(8)$.  Now gauge $G_{\rm gauge}=SU(3)$, embedded so that the Higgs field
transforms as the adjoint representation of $SU(3)$.  It is easily verified
that the most general quartic Higgs potential that is $SU(3)$ invariant  also
respects an ``accidental'' $SO(8)$ symmetry.\refmark{\pseudo}  Depending on the
alignment, the
unbroken gauge symmetry will be either $[SU(2)\times U(1)]/Z_2$ or $[U(1)\times
U(1)]/Z_2$.  Solving for the alignment by minimizing the one--loop effective
potential, one finds $H_{\rm gauge}=[SU(2)\times U(1)]/Z_2$.  Since
$\pi_2(G_{\rm gauge}/H_{\rm gauge})=Z$ and $\pi_2(G_{\rm approx}/H_{\rm
approx})=0$, this model contains semilocal monopoles.  Another example will be
described in Section 6.

\section{Unstable $Z_2$ Vortex}
The model in Section 5.1 has a spontaneously broken $U(1)_{\rm gauge}$, and the
topologically conserved magnetic flux takes integer values.  The model in this
section will demonstrate that it is also possible for unstable vortices to
occur when the topologically conserved magnetic flux takes values in $Z_2$.

The approximate global symmetry is $G_{\rm approx}=[SU(3)_L\times
SU(3)_R]/Z_3$, and the Higgs field transforms as the $(3,\bar 3)$
representation; it can be written as a $3\times 3$ matrix $\Phi$ transforming
as
$$
\Phi\to U_L\Phi U_R^\dagger~.
\eqn\noninttrans
$$
We suppose that the Higgs expectation value can be put in the form
$$
\VEV{\Phi}=v{\bf 1}~,
\eqn\nonintvev
$$
so that the pattern of symmetry breakdown is
$$
G_{\rm approx}=[SU(3)_L\times SU(3)_R]/Z_3\longrightarrow
H_{\rm approx}=SU(3)_V/Z_3~.
\eqn\notintpat
$$
(There are now two independent quartic invariants in the the most general Higgs
potential, and one cubic invariant, but this pattern occurs for a finite range
of parameters.)

Now gauge the subgroup $SO(3)\subset SU(3)_R$.  This model is natural, and the
alignment problem is trivial.  The exact symmetry breaks as
$$
G_{\rm exact}=SU(3)_L^{\rm global}\times SO(3)_R^{\rm gauge}
\longrightarrow H_{\rm exact}=SO(3)_V^{\rm global}~;
\eqn\nonintexact
$$
the gauge symmetry is completely broken.

Since $\pi_1(SO(3)_R^{\rm gauge})=Z_2$, this model has a topologically
conserved $Z_2$ magnetic flux.  But we also have $\pi_1(G_{\rm exact}/H_{\rm
exact})=0$.  (The noncontractible loop in $SO(3)_R^{\rm gauge}$ can be deformed
in $G_{\rm exact}$ to a loop that lies in $H_{\rm exact}$.)  Thus, there are
configurations with nontrivial $Z_2$ magnetic flux such that the Higgs field
lies in the exact vacuum manifold everywhere.  According to the general
discussion in Section 3, then, the vortex will be unstable when the gauge
coupling is sufficiently weak.

\chapter{Natural Models:  Dynamical Symmetry Breaking}
In some of the models described above, fine tuning of bare parameters is
required to enforce the condition that the $G_{\rm approx}$ symmetry is a good
symmetry to zeroth order in $\hbar$.  This kind of fine tuning can be avoided
in a broad class of models that contain no elementary scalar fields.  In these
models, the spontaneous breakdown of the $G_{\rm approx}$ symmetry is {\it
dynamical}, driven by the condensation of fermion pairs.

Of course, the dynamical symmetry breakdown is actually non-perturbative in
$\hbar$, rather than ``classical.''  So we need to change our terminology a
bit.  In these models, the intrinsic breaking of the $G_{\rm approx}$ symmetry
turns off as the weak $G_{\rm gauge}$ couplings go to zero.  The models are
natural in the sense that there are no operators of dimension four or less that
are invariant under $G_{\rm exact}$, other than gauge couplings.  The only
potential symmetry breaking terms are bare fermion masses, so we need to ensure
that the $G_{\rm exact}$ symmetry is sufficiently restrictive to prevent
fermion masses from being generated by the $G_{\rm gauge}$ radiative
corrections.

For example, QCD with two massless quark flavors has the chiral symmetry
$$
G_{\rm approx}=[SU(2)_L\times SU(2)_R\times U(1)_V]/Z_2~,
\eqn\qcd
$$
which is dynamically broken to
$$
H_{\rm approx}=[SU(2)_V\times U(1)_V]/Z_2~.
\eqn\qcdunbr
$$
If we now gauge $G_{\rm gauge}=[U(1)_L\times U(1)_R]/Z_2$, the surviving exact
symmetry (in fact, the gauge symmetry) is sufficient to forbid any bare quark
masses.

We may proceed to determine the vacuum alignment as in Section 5.2, but in one
important respect, the previous analysis needs to be reconsidered.  The
effective potential that we computed before was of order $e^4\ln(1/e^2)$, where
$e$ is the $G_{\rm gauge}$ gauge coupling.  But there may also be terms in the
potential that are of order $e^2$, and so are the dominant terms at weak gauge
coupling.  (We did not consider such terms before, because they are not
generated until two-loop order in models with elementary scalars.)
Fortunately, it is easy to show that no order-$e^2$ terms arise in the type of
model considered here, where no weak gauge bosons couple to both left-handed
and right-handed quarks.\Ref\peskin{M. E. Peskin, {\it Nucl. Phys.} {\bf B175}
(1980) 197; J. Preskill, {\it Nucl. Phys.} {\bf B177} (1981) 21.}  Thus, our
previous analysis of the vacuum alignment
is applicable.  The new feature is that the effective potential is actually
finite, because there are no possible symmetry-breaking counterterms; it has
the from eq.~\effpot, where $M$ is the scale of dynamical symmetry breakdown.
We conclude that the model contains a vortex and domain wall with thickness
given by eq.~\coresize.

The situation is different for our model with a semilocal monopole, in which
$G_{\rm gauge}=SU(2)_V$.  Here the exact symmetry is
$$
G_{\rm exact}=[SU(2)_V\times U(1)_V\times Z_{4,A}]/Z_2~,
\eqn\vecqcd
$$
and the axial $Z_{4,A}$ symmetry is sufficient to forbid bare quark masses.
But since the weak gauge symmetry is now vector-like, there is an order-$e^2$
term in the effective potential.  The minimum of this potential occurs when
$G_{\rm gauge}$ is unbroken,\REF\wittineq{E. Witten, {\it Phys. Rev. Lett.}
{\bf 51} (1983) 2351.}\refmark{\peskin,\wittineq} contrary to our previous
findings, and the model
contains no magnetic monopoles.

It is not difficult to construct slightly more elaborate models in which
natural semilocal monopoles can occur.  For example, the symmetry breakdown
pattern
$$
G_{\rm approx}=SU(4)\to H_{\rm approx}=Sp(4)
\eqn\spmodel
$$
is expected to occur, in a model that contains four massless fermion flavors
that transform as a pseudoreal representation of a strongly coupled gauge
group.  (Because the representation is pseudoreal, a gauge-invariant bilinear
fermion condensate must be antisymmetric in flavor indices, and $Sp(4)$ is the
maximal symmetry that preserves a condensate in which all fermions acquire
masses.)  Now, if we gauge $G_{\rm gauge}=SO(4)$ (embedded so that the 4 of
$SU(4)$ transforms as a 4 of $SO(4)$), bare fermion masses are
forbidden.  The condensate transforms as $(3,1)+(1,3)$ under $G_{\rm gauge}\sim
SO(3)\times SO(3)$, and we can use the methods of Ref.~[\peskin] to find that
the
vacuum alignment favors the gauge symmetry breakdown pattern
$$
G_{\rm gauge}=SO(4)\to H_{\rm gauge}=[SU(2)\times U(1)]/Z_2~.
\eqn\natmonopatt
$$
Since $\pi_2(G_{\rm gauge}/H_{\rm gauge})=Z$ and $\pi_2(G_{\rm approx}/H_{\rm
approx})=0$, this model contains a semilocal monopole.  Since the discrete
parity symmetry that interchanges the two $SO(3)$ factors (which is embedded in
$SU(4)$) is also spontaneously broken, there is  a semilocal domain wall in the
model, as well.

In models of dynamical symmetry breakdown, then, semilocal defects are
topological defects that can be analyzed using an effective Lagrangian that
describes physics well below the scale of the symmetry breakdown, as these
examples illustrate.  The defects
have a characteristic size that is larger than the symmetry breaking scale by a
power of the inverse $G_{\rm gauge}$ coupling.


\chapter{Mixing and Twisting}
The models that we have been considering have a
$G_{\rm exact}$ symmetry with the local structure\foot{The general compact
symmetry group with this local structure is $G_{\rm exact}=[G_1\times
G_2]/G_{\rm
discrete}$, where $G_{\rm discrete}$ is a discrete invariant subgroup of
$G_1\times G_2$.  But there is really no loss of generality in assuming $G_{\rm
exact}=G_1\times G_2$, if we allow matter fields that represent
$G_{\rm discrete}$  trivially.}
$$
G_{\rm exact}\sim G_1\times G_2~,
\eqn\localgroup
$$
where $G_1$ is the gauge group and $G_2$ is a global symmetry group.  In
Section 3, we considered the properties of semilocal vortices in models such
that $G_2$ is a nontrivial continuous group.  We saw that, under suitable
conditions, there can be topological magnetic flux
sectors that contain configurations of finite energy in which the flux is
spread out over an arbitrarily large area.  Such configurations exist if the
Higgs field on the circle at $r=\infty$ traces out a noncontractible path in
$G_1/H_1$, and this path can be contracted in $[G_1\times G_2]/H$---in other
words, if the vortex is classified by an element of the kernel of the natural
homomorphism
$$
\pi_1(G_1/H_1)\to \pi_1([G_1\times G_2]/H)~.
\eqn\electrohomo
$$
In this section, I will discuss this criterion in a bit more detail.
Specifically, I will emphasize the (rather obvious) fact that the kernel can be
nontrivial only if ``mixing'' occurs;  that is, there must be a generator of
$H$ that is a nontrivial linear combination of a $G_1$ generator and a $G_2$
generator.

To see this, let us recall that a closed
loop in the coset space $G/H$ may be expressed as
$$
\eqalign{\Phi(\theta)&=D[g(\theta)]\Phi_0~,~~g(\theta)\in G~,\cr
&g(0)=e~,\quad g(2\pi)\in H~;\cr}\eqn\orbit
$$
here $\Phi$ is an ``order parameter'' with stability group $H$, and $D$ is the
representation of $G$ according to which $\Phi$ transforms.
Thus, closed paths in $G/H$ that begin and end at
an arbitrarily selected point $\Phi_0$ are parametrized by paths in $G$ (open,
in general), that begin at the identity and end at a point in
$H$.  The homotopy classes in $\pi_1(G/H)$,
then, are equivalent to topological classes of paths in $G$ that
begin at the identity and end in $H$.  There are two types of
nontrivial classes---ones that end in the identity component of $H$
(which occur only if $G$ is not simply connected), and ones that
don't (which occur only if $H$ is not connected).

In the case $G=G_1\times G_2$, a closed path in $[G_1\times G_2]/H$
can be expressed as
$$
\eqalign{\Phi(\theta)&=D[g_1g_2(\theta)]\Phi_0~,\quad
g_1(\theta)\in G_1~,\quad g_2(\theta)\in G_2~,\cr
&g_1g_2(0)= e~,\quad g_1g_2(2\pi)\in H~,\cr}
\eqn\electroloop
$$
where $D$ is the representation of $G_1\times G_2$ according to which $\Phi$
transforms.  Now, consider a nontrivial element of the kernel of the
homomorphism eq.~\electrohomo.  Representing it is a path $g_1^{(0)}(\theta)\in
G_1$ that cannot be
smoothly deformed so that it lies in $H_1$ for all $\theta$, if we fix
$g_1(0)=e_1$, and require that $g_1(2\pi)\in H_1$.  By assumption, it is
possible to deform this path so that $g_1g_2(\theta)$ lies entirely in
$H$.

Let us denote this deformation by $g_1g_2(t,\theta)$, where $t\in [0,1]$, and
$$
\eqalign{&g_1(0,\theta)=g_1^{(0)}(\theta)~,\quad g_2(0,\theta)=e_2~,\cr
&g_1g_2(1,\theta)\in H~.\cr}
\eqn\deform
$$
Now we distinguish two cases.  If $g_1(t,2\pi)\in H_1$ for all $t$, then we
know that $g_1(1, \theta)$ cannot lie in $H_1$ (for otherwise
$g_1^{(0)}(\theta)$ defines a trivial closed path in $G_1/H_1$, contrary to our
assumption).   But $g_1(1,\theta)g_2(1,\theta)$ {\it is} in $H$.  So, as
$\theta$ varies, $D[g_1(1,\theta)]$ and $D[g_2(1,\theta)]$ both act
nontrivially on  the order parameter $\Phi_0$, while their product acts
trivially.  This means that there is a generator of $H$ that is a nontrivial
linear combination of broken $G_1$ and $G_2$ generators---in other words, there
is mixing.

On the other hand, suppose that $g_1(t,2\pi)$ does not stay in $H_1$ for all
$t$.  Then, since $g_1(t,2\pi)g_2(t,2\pi)\in H$, we know that, as t varies,
$D[g_1(t,2\pi)]$ and $D[g_2(t, 2\pi)]$ act nontrivially on $\Phi_0$, while
their product acts trivially.  Again, we conclude that there is mixing.

It is useful to restate this conclusion in the language of fiber bundles.
We noted in Section 2 that the Nambu--Goldstone bosons associated with the
vacuum manifold $[G_1\times G_2]/H$ can be divided into two classes---those
that are eaten by the $G_1$ gauge fields and the surviving Nambu--Goldstone
bosons that remain exactly massless.  This division defines, locally, a
decomposition of the vacuum manifold into a direct product of two spaces---the
$G_1$ gauge orbit and the space $M_{\rm orbit}$ of gauge orbits.  In other
words, there is a projection map
$$
\pi~: [G_1\times G_2]/H\longrightarrow M_{\rm orbit}
\eqn\bundleproj
$$
that takes each point of the vacuum manifold to the gauge orbit on which it
lies.  This map is a {\it fibration} of the vacuum manifold, with base space
$M_{\rm orbit}$, fiber $G_1/H_1$ (the gauge orbit), and structure group $G_1$.

Now, the topologically conserved magnetic flux is classified by the fundamental
group of the fiber, the gauge orbit.  Configurations with nontrivial magnetic
flux can ``spread out'' if there are noncontractible loops in the fiber that
can be contracted in the total space of the bundle---that is, if the
homomorphism eq.~\electrohomo\ has a nontrivial kernel.

But suppose that there is no mixing---the unbroken group is $H=H_1\times H_2$,
where $H_1\subseteq G_1$ and $H_2\subseteq G_2$.  Then we have
$$
{G_1\times G_2\over H_1\times H_2}
={G_1\over H_1}\times {G_2\over H_2}~;
\eqn\nonmixfactor
$$
the vacuum manifold is  {\it globally} a direct product of the gauge orbit
$G_1/H_1$ and the space $M_{\rm orbit}=G_2/H_2$.  Thus, noncontractible loops
in a gauge orbit evidently remain noncontractible in the total space of the
bundle.  Vortices with nontrivial magnetic flux cannot spread.

For a vortex to be able to spread, it is necessary (but not sufficient) for the
the vacuum bundle to be a nontrivial (``twisted'') bundle with base space
$M_{\rm orbit}$ and fiber $G_1/H_1$.  For the bundle to be twisted, it is
necessary (but not sufficient) for mixing to occur.

Magnetic monopoles are classified by noncontractible two-spheres in the gauge
orbit.  As noted in Section 4.2, such a two-sphere always remains
noncontractible in the total space of the bundle.  A magnetic monopole (with
nontrivial topological charge) always has a core.

\chapter{(Generalized) Electroweak Vortices}
As noted above, in a magnetic flux sector classified by a nontrivial element of
the kernel of the homomorphism eq.~\electrohomo, there are
configurations of finite energy in which the flux is
spread out over an arbitrarily large area.  It then becomes a dynamical
question whether the energy is minimized in this sector by a spread out
configuration or a localized vortex.  We argued in Section 3 that the spread
out
configurations are favored at sufficiently weak gauge coupling, but that stable
localized vortices may exist if the gauge coupling is not too weak (or the
Higgs mass is not too large).

Following Vachaspati,\refmark{\vach} let us consider what would happen to such
a stable vortex
if we were to gauge the global $G_2$ symmetry.  When $G_2$ is gauged, the
vortex no longer carries a topologically conserved magnetic flux, so it is
bound to become unstable.  But we know that the vortex is stable in the limit
$e_2\to 0$, where $e_2$ is the $G_2$ gauge coupling.  It is reasonable to
expect that the classical vortex solution remains classically stable for a
finite range of values of $e_2$, though there are presumably quantum mechanical
tunneling processes that allow it to decay.  As Vachaspati observes, if we
gauge the $SU(2)_L$ global symmetry in the model described in Section 5.1, we
obtain the standard electroweak model.  This model therefore contains
metastable ``electroweak strings,'' (although not for realistic values of the
Higgs mass and $\sin^2\theta_W$\Ref\perivach{M. James, L. Perivolaropoulos and
T. Vachaspati, private communication.}).

In this Section, I will discuss a few features of the theory of such
electroweak  vortices.

In general, we consider a model with gauge group $G_1\times G_2$, spontaneously
broken to $H$.  If the $G_2$ gauge coupling $e_2$ turns off, the gauge group
$G_1$ breaks to $H_1$, the intersection of $G_1$ and $H$.  A (generalized)
electroweak vortex is a vortex that carries no topologically conserved flux,
but becomes topologically stable in the limit $e_2\to 0$; thus, it is
associated with an nontrivial element of the kernel of eq.~\electrohomo.  As is
clear from the discussion in Section 7, such an object can exist only if there
is gauge boson mixing---there must be a generator of $H$ that is a nontrivial
linear combination of a $G_1$ generator and a $G_2$ generator.

\section{Strings Ending on Monopoles}
Let us denote by $Q_{1,2}$ two generators of $G_{1,2}$ that mix.  Suppose that
the Higgs field $\Phi^{q_1,q_2}$ carries charges $q_{1,2}$, so that
$$
Q={Q_2\over q_2}-{Q_1\over q_1}
\eqn\mixunbroken
$$
is an unbroken $H$ generator.  If $B_{1,2}$ are the gauge fields that couple to
$Q_{1,2}$, then
$$
A=~B_1\cos\theta +B_2\sin\theta
\eqn\mixmatrix
$$
is the massless gauge field that couples to $eQ$, where $e$ is related
to the $G_{1,2}$ gauge couplings by
$$
{e\over\sin\theta}=e_2q_2~,\quad {e\over \cos\theta}=-e_1q_1~.
\eqn\thetadef
$$
The orthogonal gauge field state is
$$
Z=-B_1\sin\theta +B_2\cos\theta~,
\eqn\mixZ
$$
which couples to
$$
e_ZQ_Z={e\over\cos\theta\sin\theta}\left({Q_1\over q_1}\sin^2\theta
+{Q_2\over q_2}\cos^2\theta\right)~=~
{e\over\cos\theta\sin\theta}\left({Q_2\over q_2}-Q\sin^2\theta\right)~.
\eqn\Zcurrent
$$
The $Z$ need not be a mass eigenstate field; it could be a linear combination
of massive gauge bosons with different masses.  For example, we might have
$Z=X\cos\tilde\theta+Y\sin\tilde\theta$, where $X$ is a mass eigenstate
coupling to $e_XQ_X$ and $Y$ is a mass eigenstate coupling to $e_YQ_Y$.  Then
eq.~\Zcurrent\ is the combination
$e_XQ_X\cos\tilde\theta+e_YQ_Y\sin\tilde\theta$ that couples to $Z$.
(Note also that $G_1$ or $G_2$ could  be a product of several
commuting factors, each with an independent gauge coupling.
Then $Q_1$, for example, might  be a linear combination of generators, each
belonging to a different invariant subalgebra of the $G_1$ Lie algebra.)

Now consider a vortex that has $Z$ magnetic flux $\Psi^{(Z)}$ confined to its
core.   This means that, at least in a particular gauge, we have
$$
P\exp\left(i\oint_C e_a Q^a B^a_\mu dx^\mu\right)
=\exp\left(i e_Z Q_Z \Psi^{(Z)}\right)~,
\eqn\fluxdefine
$$
where $C$ is a closed path that encloses the vortex.  Here  $B_\mu^a$ has been
summed over the $G_1\times G_2$ gauge fields, and $e_a$, $Q^a$ are the
corresponding gauge couplings and generators.
Since
the Higgs field $\Phi^{q_1,q_2}$ must be covariantly constant and single-valued
outside the core, the $Z$ flux is required to be an integer multiple of the
flux quantum
$$
\Psi^{(Z)}_0={2\pi\over e}\cos\theta\sin\theta~=~
-{2\pi \sin\theta\over e_1q_1}~.
\eqn\fluxquan
$$
If a particle is covariantly transported around the minimal vortex, it acquires
the Aharonov--Bohm phase
$$
\exp\left[2\pi i\left({Q_2\over q_2}-Q\sin^2\theta\right)\right]~.
\eqn\ABphase
$$
For a typical charged particle, and a generic value of the mixing angle
$\theta$, this is a nontrivial (in fact, transcendental) phase.  But it follows
from our assumption that the vortex carries no conserved topological charge
that eq.~\ABphase\ is an element of the  identity component of $H$.  In two
spatial dimensions, this means that it is possible to smoothly deform the
vortex configuration (while the energy remains finite) to a configuration that
has only massless $H$ magnetic flux.  This configuration can then lower its
energy to zero by spreading out indefinitely.  Thus, though the vortex may be
classically stable, it can decay by tunneling quantum mechanically to the
configuration with massless magnetic flux.

Similarly, in three spatial dimensions, there are configurations in which the
$Z$ flux tube ends on a finite-mass ``magnetic monopole,'' with  $A$ magnetic
flux spilling out of the end.  One may regard the flux tube as a {\it visible}
Dirac string; then the magnetic flux through a sphere enclosing the monopole
may be inferred from eq.~\ABphase.
Let us define the magnetic charge $g_{\rm mag}$ of the monopole so that $4\pi
g_{\rm mag}$ is the total
$A$ magnetic flux emanating from the monopole;  more precisely, let
$$
P\exp \left(i\oint_C e Q A^\mu dx_\mu\right)=\exp \left(4\pi i e Q g_{\rm
mag}\right)~,
\eqn\magchardefine
$$
where $C$ is a path that encloses the Dirac string of the monopole.  We
conclude that
$$
g_{\rm mag}={1\over 2e} \sin^2\theta~.
\eqn\nambuflux
$$
Of course, this flux does not satisfy the Dirac quantization condition, because
the string is not invisible.  (If $\exp(2\pi iQ_2/q_2)$ is a nontrivial element
 of the identity component of $H$, there will be some additional magnetic flux,
coupling to another $H$ generator, aside from the $A$ flux given by
eq.~\nambuflux.)  Note that, in the case of the standard model, $Q$ defined by
eq.~\mixunbroken\ is actually twice the conventionally normalized electric
charge operator, so we have $g_{\rm mag}=\sin^2\theta/e$, if $e$ is the
conventionally normalized electromagnetic gauge coupling.  Such $Z^0$ magnetic
flux tubes ending on magnetic monopoles were first discussed by
Nambu.\Ref\nambu{Y. Nambu, {\it Nucl. Phys.} {\bf B130} (1977) 505.}

A classically stable electroweak string can break in a quantum mechanical
tunneling process where a pair of monopoles nucleates spontaneously.  The decay
of metastable electroweak vortices and flux tubes will be further discussed in
Ref.~[\presvile].

\section{Aharonov--Bohm Interactions}
We have seen that, for generic values of the mixing angle, particles with
nonvanishing $Q$ have nontrivial Aharonov--Bohm interactions with electroweak
strings.  In principle, the charge $Q$ of a projectile could be measured by
scattering the projectile off of a string.

Such measurement processes have attracted much recent interest, particularly in
the case where the unbroken gauge group $H_{\rm gauge}$ is
disconnected.\Ref\krauss{L. M. Krauss and F. Wilczek, {\it Phys. Rev. Lett.}
{\bf 62} (1989) 1221.}  In
that case, there are topologically stable strings associated with the ``local
discrete symmetry.''  The Aharonov--Bohm interaction can then probe the
``quantum hair'' of an object.  This quantum hair can be measured at long
range, but becomes invisible in the classical limit.  In the case of an
electroweak string, however, the flux of the string is in the identity
component of $H_{\rm gauge}$, and the string is not topologically stable.  The
charges that can be measured in Aharonov--Bohm scattering off the string are
not varieties of quantum hair.
To be specific, consider the standard model, in which $\exp(2\pi iQ_2/q_2)=1$.
Then the Aharonov-Bohm phase \ABphase\ is completely determined by the charge
$Q$.
Therefore, we can not learn anything about a particle in an Aharonov-Bohm
scattering experiment that we could not discern by measuring its classical
electric field.

This observation is easily generalized.  The effect of transport around a
vortex is always described by an element of the unbroken gauge group $H_{\rm
gauge}$, because the Higgs condensate must be covariantly constant and single
valued outside the vortex.  Thus, the Aharonov-Bohm phase acquired by any
projectile is always determined by its transformation properties under $H_{\rm
gauge}$.   The ``classical hair'' of the projectile determines its charges in
the $H_{\rm gauge}$ Lie algebra.  This leaves undetermined only the
transformation properties under the ``local discrete symmetries'' that are not
in the identity component of $H_{\rm gauge}$.  These additional charges cannot
be measured in Aharonov--Bohm scattering if the flux of the string is in the
identity component.  Thus, quantum hair can be measured only with topologically
stable strings.


\section{Embedded Defects}
Vachaspati and Barriola\refmark{\vachbarr} have recently pointed out a more
general procedure
for constructing
static solutions to the classical field equations that are not topologically
stable.  Consider a gauge theory with gauge group $G_{\rm gauge}$ spontaneously
broken to the subgroup $H_{\rm gauge}$.  Now choose a nontrivial subgroup
$\tilde G_{\rm gauge}$, such that the intersection of $\tilde G_{\rm gauge}$
and $H_{\rm gauge}$ is $\tilde H_{\rm gauge}$.  Suppose that the natural
homomorphism
$$
\pi_n(\tilde G_{\rm gauge}/\tilde H_{\rm gauge})\longrightarrow
\pi_n(G_{\rm gauge}/H_{\rm gauge})
\eqn\embedhomo
$$
has a nontrivial kernel.  In other words, there are noncontractible loops (n=1)
or two-spheres (n=2) in $\tilde G_{\rm gauge}/\tilde H_{\rm gauge}$ that are
contractible in $G_{\rm gauge}/H_{\rm gauge}$.  Vachaspati and Barriola then
show that, under suitable conditions,
there are classical vortex  (n=1) or monopole (n=2) solutions to the field
equations associated with the nontrivial elements of the kernel.  That is, if a
gauge theory with gauge group $\tilde G_{\rm gauge}$ broken to $\tilde H_{\rm
gauge}$ contains a topologically stable defect, this defect remains a solution
to the field equations when the gauge group is enlarged to $G_{\rm
gauge}\supset \tilde G_{\rm gauge}$.  The electroweak vortices described above
are a special case of such ``embedded defects,'' where $G_{\rm gauge}=G_1\times
G_2$ and $\tilde G_{\rm gauge}=G_1$.

But there is no particular reason, in general, to expect an embedded defect to
be classically stable.  In the case of embedded monopoles, one can make a
stronger statement:  they are always classically {\it unstable} (if not
topologically stable).

To see this, we recall that, in a model with unbroken $H_{\rm gauge}$ symmetry,
we may associate with any magnetic monopole a topological $H_{\rm gauge}$
charge.  The matching condition (or Dirac string) of the monopole defines a
closed path in $H_{\rm gauge}$, and the corresponding element of $\pi_1(H_{\rm
gauge})$ is the magnetic charge.\refmark{\lubkin,\coleman}   If the monopole
arises in a model with an
underlying $G_{\rm gauge}$ symmetry, and is nonsingular, then this loop in
$H_{\rm gauge}$ must be contractible in $G_{\rm gauge}$.  (Otherwise, the Dirac
string would necessarily end on a point singularity.) This means that a
nonsingular monopole with nontrivial $H_{\rm gauge}$ is always associated with
a nontrivial element of $\pi_2(G_{\rm gauge}/H_{\rm gauge})$---it is
topologically stable.

Conversely, a monopole that is {\it not} topologically stable must carry
trivial $H_{\rm gauge}$ charge.   It was shown by Brandt and Neri\Ref\neri{R.
Brandt and F. Neri, {\it Nucl. Phys.} {\bf B161} (1979) 253.} and
Coleman\Ref\coledecay{S. Coleman, ``The Magnetic Monopole Fifty Years Later,''
in {\it The Unity of the Fundamental Interactions}, ed. A. Zichichi (Plenum,
London, 1983).}
that such monopoles are always classically unstable.  To demonstrate the
instability, it suffices to study the small vibrations of the long range
$H_{\rm gauge}$ gauge field; it is not necessary to consider the structure of
the monopole core.  But since there is no topological conservation law to
prevent it, the core will presumably ``unwind,'' and its energy will be carried
to spatial infinity as non-abelian radiation.

An embedded monopole is just a particular type of monopole solution that
carries no topological charge, and it is therefore unstable.

\chapter{Concluding Remarks}
\section{Semilocality}
I have used the term ``semilocal'' to characterize defects that occur in models
in which the gauge group is embedded in a larger group of (approximate) global
symmetries.
These defects carry ``topologically conserved'' charges, yet can be deformed so
that the order parameter takes values in the approximate vacuum manifold
everywhere.
This usage encompasses the vortices originally considered by Vachaspati and
Ach\'ucarro.\refmark{\vachachu}  It also includes a broader class of domain
walls, vortices, and monopoles.  These share the feature that the spontaneously
broken approximate global symmetry is not restored inside the core of the
defect.  Indeed, the structure of the defect can be well described using an
effective field theory, in which the physics responsible for the spontaneous
symmetry breakdown has been ``integrated out.''

But the term ``semilocal'' could be and has been used in other ways.
Hindmarsh\Ref\hindagain{M. Hindmarsh, ``Semilocal Topological Defects,'' DAMTP
Preprint DAMTP-HEP-92-24 (1992).} defines a semilocal defect as one that arises
in a model such that the vacuum manifold is a twisted bundle of gauge orbits,
as described in Section 7.  This classification leads him to consider an
interesting ``semilocal texture'' contained in the model of Ref.~\vachachu\ and
Section 5.1.  I have not discussed textures in this paper, as the emphasis has
been on defects that carry topologically conserved charges.  (Textures, in
contrast, even if classically stable, can ``unwind'' via quantum tunneling.)

\section{Semilocal Strings}
Although I have broadened the notion of a semilocal defect here, it should be
noted that the most interesting kind of semilocal defect is still the semilocal
vortex analyzed in Ref.~\vachachu.  I would like to emphasize what was truly
surprising and noteworthy (to me) about Ref.~\vachachu.  It was not, I think,
that a stable vortex could exist even though the vacuum manifold is simply
connected.  It had been stressed by Coleman,\foot{See
Appendix 3 of Ref.~[\coleman].} and was widely appreciated, that only the
pattern of {\it gauge} symmetry breaking is relevant to the classification of
finite--energy vortices.  The surprise was not that a semilocal vortex could be
{\it stable}, but that it could be {\it unstable}.  Part of the motivation for
this work came from the desire to understand better why the magnetic flux wants
to spread out when the gauge coupling is sufficiently weak.  (It is also nicely
explained in Hindmarsh's papers.\refmark{\hindmarsh,\hindagain})

\section{Electroweak Strings}
Having said that the surprising feature of semilocal vortices is that they can
decay, I should admit that the implications of the existence of {\it stable}
semilocal vortices are quite interesting.  As Vachaspati\refmark{\vach}
emphasized, a stable semilocal vortex will remain {\it classically} stable even
if the global symmetry is gauged, provided the gauge coupling is not too large.
Unfortunately, classically stable strings do not arise in the minimal standard
model, for realistic values of the $\sin^2\theta_W$ and Higgs
mass.\refmark{\perivach}  But they may well occur in realistic extensions of
the standard model.  Thus, we are invited to contemplate the consequences of
long--lived metastable strings at the electroweak scale.

First, there would be  new resonances at the TeV scale.  These could be
segments of string with monopoles at the ends (as envisioned long ago by
Nambu\refmark{\nambu}), or closed loops of strings.  Regrettably, since these
states are ``squishy'' classical objects, production of the new resonances
would be highly suppressed in hard pointlike collisions.  They are not likely
to be seen in future accelerator experiments.

Second, the strings would be produced during the electroweak phase transition
in the early universe.  Not many strings would survive to the present epoch,
though.  Because the strings can end on monopoles, the strings that are
initially produced in the phase transition will be predominantly short open
segments and small closed loops.
\Ref\mitchell{D. Mitchell and N. Turok, {\it Nucl. Phys.} {\bf B294} (1987)
1138; D. Haws, E. Copeland, T. W. B. Kibble, D. Mitchell, and N. Turok, {\it
Nucl. Phys.} {\bf B298} (1988) 445; A. Everett, T. Vachaspati, and A. Vilenkin,
{\it Phys. Rev.} {\bf D31} (1985) 1925.}
Crudely speaking, each string has a nonzero probability per unit length of
ending (on a monopole), so that long strings are exponentially suppressed.  The
string--monopole network is therefore expected to disappear quickly.  The main
cosmological implications of the strings, then, would concern their influence
on the electroweak phase transition itself, perhaps including their impact on
electroweak baryogenesis.\Ref\davis{R. H. Brandenberger and A.-C. Davis,
``Electroweak Baryogenesis and Electroweak Strings,'' Brown Preprint
BROWN-HET-862 (1992).}

\section{Electroweak Flux Tubes and the Monopole Problem}
Another potential cosmological implication of electroweak strings deserves
comment.  Lazarides and Shafi\Ref\shafi{G. Lazarides and Q. Shafi, {\it Phys.
Lett.} {\bf 94B} (1980) 149.} suggested many years ago that electroweak flux
tubes might offer a natural solution to the cosmological monopole
problem.\Ref\presmono{J. Preskill, {\it Phys. Rev. Lett.} {\bf 43} (1979)
1365.}  The idea is that the GUT monopoles that are copiously produced in the
very early universe might become confined by flux tubes after the electroweak
phase transition.  The flux tubes would greatly enhance the rate of monopole
annihilation, and rapidly reduce the monopole abundance to an acceptable level.

There are some problems with this idea.  First, as Lazarides and Shafi
noted,\refmark{\shafi} the magnetic monopoles in the simplest grand unified
models carry $U(1)_{\rm electromagnetic}$ magnetic charge and $SU(3)_{\rm
color}$ magnetic charge.  They do not have any $Z^0$ magnetic flux, and they
are little affected by the electroweak phase transition.  Still, there are
alternative models in which the stable magnetic monopoles carry $U(1)_{\rm
hypercharge}$ magnetic charge (as well as color magnetic charge).  These
monopoles have both $Z^0$ and $A$ magnetic flux, so that the Lazarides--Shafi
mechanism might work.

A second problem is that the $Z^0$ flux tubes are unstable in the simplest
models, so that monopole confinement does not really occur, even if the
monopoles do have $Z^0$ magnetic fields.  But we have noted that the $Z^0$ flux
tubes could be stable in extended models, so it still seems that there is a
class of models in which the Lazarides--Shafi mechanism could work.

There is a third problem however, that probably makes the idea untenable, even
under optimistic assumptions.  The problem is that an electroweak flux tube can
end on either a heavy GUT monopole or on a light electroweak (Nambu) monopole.
There is no guarantee, then,  that the flux tube emanating from a GUT monopole
will bind it to another GUT monopole, rather than to a light electroweak
monopole.

The GUT monopole with minimal $U(1)_{\rm hypercharge}$ magnetic charge carries
electromagnetic magnetic charge $\cos^2\theta/e$, in addition to its confined
$Z^0$ flux.  If the flux tube ends on an electroweak monopole with charge
$\sin^2\theta/e$, then the monopole-string composite has magnetic charge $1/e$,
twice the Dirac charge.  After the flux tube shrinks away, this object becomes
an unconfined stable magnetic monopole, with electromagnetic (and color)
magnetic charge.

For the Lazarides--Shafi mechanism to successfully reduce the monopole
abundance to an acceptable level, electroweak monopoles must be heavily
suppressed, so that the flux tubes almost always end on GUT monopoles.  It
seems difficult to devise a plausible scenario of this kind.

\bigskip
As this paper was being completed, I became aware of Ref.~\hindagain, which has
some overlap with the research reported here.
\ack
I thank Tanmay Vachaspati for discussions that stimulated my interest in this
subject.  I have also benefitted from enjoyable discussions with Robert
Brandenberger, Martin Bucher, Steve Frautschi, Mark Hindmarsh, and Hoi-Kwong
Lo.

\refout
\bye